**Multiple Instrument Methods Comparison by Precision-weighted Deming Regression**

Douglas M Hawkins,[1]

Jessica J Kraker[2]

School of Statistics, University of Minnesota, Ford Hall, Minneapolis MN 55455, USA
dhawkins@umn.edu.

Mathematics Department, University of Wisconsin, Eau Claire, WI, USA.

**Abstract**

In methods comparison (MC) studies, specimens are tested using two or more instruments with the objective of establishing the statistical relationship between the different instruments' readings. Unlike regular regression, this is an errors in variables problem. Relationships may be fitted parametrically ("Deming" regression) or non-parametrically ("Passing Bablok or PB" regression.) In clinical chemistry settings, the measurement variability is rarely constant, but generally increases with increasing analyte values. Precision weighted Deming regression models this variability and incorporates it into the fitting. The simplest setting of comparing two instruments is discussed in (1) and implemented in an R package (2). PB makes minimal distributional assumptions. Its classical two-instrument implementation has recently been extended to multiple instruments (3). This work extends the two-instrument Deming model of (1) to multiple instruments, developing algorithms for fitting, for formal inference, for residual analysis, and for outlier detection and diagnosis.

## Introduction

Suppose $n$ samples are measured using $I$ instruments, giving a matrix X of readings. Write $\mu$ for the latent true concentrations of the samples. A possible distributional model extending (1) is

$$X_{ji} \sim N[\alpha_i + \beta_i\mu_j, \lambda_i g(\alpha_i + \beta_i\mu_j)], j=1, 2, \ldots n, \; i = 1, 2, \ldots I \qquad (1)$$

This assumes that all instruments' precision profiles have the same shape $g$, but with possibly-different scale multipliers $\lambda$.

The $\mu$ are defined only up to an arbitrary linear transformation. If one instrument (assume it is the first) is a comparator for all others, then we can resolve the indeterminacy by setting $\alpha_I = 0$, $\beta_I = 1$. Conversely, for an egalitarian analysis of the instruments, we can define a "consensus" $\mu$ by normalizing the intercepts to average zero and the slopes to average 1. This is the approach followed here.

The -2 log likelihood of the data set is:

$$L = \sum_{i=1}^{I} \sum_{j=1}^{n} \left[ \frac{(X_{ji} - \alpha_i - \beta_i\mu_j)^2}{\lambda_i g(\alpha_i + \beta_i\mu_j)} + \log[\lambda_i g(\alpha_i + \beta_i\mu_j)] \right] \qquad (2)$$

or, abbreviating $g_{ij}$ for $g(\alpha_i + \beta_i\mu_j)$

$$L = \sum_{i=1}^{I} \sum_{j=1}^{n} \left[ \frac{(X_{ji} - \alpha_i - \beta_i\mu_j)^2}{\lambda_i g_{ji}} + \log(\lambda_i g_{ji}) \right]$$

The $\alpha, \beta$ coefficients are given by weighted least squares regressions of the $X$ on $\mu$ with weights $1/g$: a separate regression for each instrument. Ignoring the second-order contribution from the log(g) terms and differentiating the quadratic term with respect to $\mu_j$ gives

$$\frac{\partial L}{\partial \mu_j} = \sum_{i=1}^{I} \frac{-2\beta_i(X_{ji} - \alpha_i - \beta_i\mu_j)}{\lambda_i g_{ji}} = \sum_{i=1}^{I} \frac{-2\beta_i(X_{ji} - \alpha_i)}{\lambda_i g_{ji}} + \mu_j \sum_{i=1}^{I} \frac{2\beta_i^2}{\lambda_i g_{ji}}$$

setting which to zero gives the maximum likelihood estimate (MLE) of $\mu_j$ as

$$\hat{\mu}_j = \sum_{i=1}^{I} \frac{\beta_i(X_{ji} - \alpha_i)}{\lambda_i g_{ji}} / \sum_{i=1}^{I} \frac{\beta_i^2}{\lambda_i g_{ji}} \tag{3}$$

Here, and later, we distinguish an estimate from its true parameter value by the statistical convention of a caret, so $\hat{\mu}$ is the estimate of $\mu$ and similarly for the other Greek letters. As the estimates of $\mu$ depend on those of $\alpha, \beta$ and vice versa, these calculations lead to alternating optimization: at each outer iteration, the estimates of $\alpha, \beta$ are computed using the current estimates of $\mu$ and recentered so that they average 0 and 1 respectively, and the $\mu$ estimates are recomputed using the fresh $\alpha, \beta$ estimates, iterating until the $\hat{\mu}$ stabilize.

Specializing the precision profile to the Rocke-Lorenzato precision profile model (1)

$$g(\mu) = \sigma^2 + \kappa^2\mu^2 = \kappa^2(\rho^2 + \mu^2), \quad \text{and defining } \rho = \sigma / \kappa,$$

$$L = \sum_{i=1}^{I} \sum_{j=1}^{n} \left[ \frac{(X_{ji} - \alpha_i - \beta_i\mu_j)^2}{\lambda_i\kappa^2\{\rho^2 + (\alpha_i + \beta_i\mu_j)^2\}} + \log[\lambda_i\kappa^2\{\rho^2 + (\alpha_i + \beta_i\mu_j)^2\}] \right] \tag{5}$$

This equation shows the parameters to be estimated: the single $\kappa$ and $\rho$, the $I$ coefficients $\alpha, \beta, \lambda$, the $n$ latent true concentrations $\mu$. Write

$$V_i = \sum_{j=1}^{n} \frac{(X_{ji} - \alpha_i - \beta_i\mu_j)^2}{\rho^2 + (\alpha_i + \beta_i\mu_j)^2}$$

$$B = \sum_{i=1}^{I} \sum_{j=1}^{n} \left[ \log[\lambda_i\{\rho^2 + (\alpha_i + \beta_i\mu_j)^2\}] \right]$$

$$L = \sum_{i=1}^{I} \frac{V_i}{\lambda_i\kappa^2} + nI\log(\kappa^2) + B \tag{6}$$

The MLE of $\kappa$ is $\hat{\kappa}^2 = \sum_{i=1}^{I} V_i / (\lambda_i nI)$ , so L is

$$L = nI + nI\log(\sum_{i=1}^{I} V_i / \lambda_i nI) + B = nI\{1 - \log(nI)\} + \log(\sum_{i=1}^{I} V_i / \lambda_i) + B$$

The expectation of $V_i$ is $n\lambda_i\kappa^2$, so $V_i / (n\kappa^2)$ estimates $\lambda_i$.

This disposes of all parameters except the Rocke-Lorenzato constant $\rho$ which can be estimated by univariate optimization. Standard errors for all parameters can be found by jackknifing.

A complication is that the likelihood is degenerate – it becomes infinite if any $\lambda$ goes to zero, so care is needed in the fitting. The procedure used for fitting the model is:

- Set all $\lambda$ to 1 and estimate all parameters.

If the true $\lambda$ differ, their estimates will be biased. If one $\lambda$ is much smaller than the others its estimate will be biased upwards because that instrument, which should be overweighted in the estimated $\mu$, will have the same weight as the others, magnifying its residuals and leading to an unduly large estimate of its $\lambda$. On the other side, if one $\lambda$ is much larger than the others, that instrument will have full weight in the estimates of $\mu$, rather than the appropriate smaller weight, shrinking its residuals and underestimating its $\lambda$. Combining these, the estimated $\lambda$, large and small, will be shrunk towards 1.

This motivates a refinement step:

- Set the $\lambda$ to the estimates produced in the first step, and re-estimate all parameters.

One might think to iterate this refinement until it converges, but its convergence is generally toward a degenerate solution with one of the $\lambda$ at zero. For this reason, it is inadvisable to go beyond a single refinement step.

Figures 1-3 are illustrative. They summarize 10,000 data sets of size n=120 with four instruments, true $\mu$ in geometric progression from 8 to 80, $\sigma$=2, $\kappa$=0.08, $\alpha$=(1, −1, 2, −2), and $\beta$=(0.9, 1.1, 1.2, 0.8)

The $\lambda$ values are 2.56, 0.16, 0.64 and 0.64, with a sixteenfold difference from first to second instruments, the last two being intermediate. These choices give a moderately challenging departure from the "null" case $\alpha$=0, $\beta$=$\lambda$=1.

The figures are comparative box and whisker plots based on breaking the data set by its abscissa into 20 equal-frequency bins. Each box describes one of these bins. Its edges are the quartiles of the abscissa and ordinate in the bin, and the interior horizontal line is at the median of the ordinate. The interior vertical line is at the median of the abscissa and extends across the range of the ordinate. A lowess smooth is shown as a dotted curve. The dotted and dashed vertical lines show the true value of the parameter and the mean of the unrefined initial estimate. The dotted and dashed horizontal lines show the true value of the parameter and the mean of the estimates after one step of refinement.

Figure 1 displays the estimates of $\lambda$ for the four instruments and confirms the substantial bias in $\lambda$ without refinement. The vertical reference lines show that the true values of $\lambda_1$ and $\lambda_2$ do not even lie within the range covered by the 10,000 data sets. The horizontal reference lines though are quite benign. For $\lambda_3$ the refinement is a wash, but it leads to a substantial bias reduction for the other three instruments.

Even better news is seen in the estimated intercepts and slopes in Figures 2 and 3. There is not much bias in the intercepts and slopes even without refinement, and one step of refinement is enough to make the bias imperceptible. Simulations exploring a variety of $\lambda$ vectors have corroborated this pattern – bias in the unrefined $\lambda$ estimates when they differ, with a substantial reduction in bias after one refinement step. The regression slope and intercept have little bias without refinement, and even less after one step.

This robustness of the regression parameters is not too surprising. Using the wrong $\lambda$ values perturbs the estimated $\mu$, but in realistic MC data sets any such perturbation is no more than a second order effect, as the variation of the true $\mu$ from sample to sample swamps the minor perturbations caused by changing the instruments' weights in computing the $\mu$ estimates. The recommended approach is therefore to use a single refinement step.

A simulation to explore this generated 10,000 data sets with four instruments, $n$=120 and true values in geometric progression from 8 to 80. The four $\alpha$ values were normal with mean vector (0,1,2,3) and standard deviation 2; the $\beta$ values were normal with mean vector (0.7, 0.9, 1.1, 1.3) and standard deviation 0.1; and the $\lambda$ values were (0, 0.2, 0.4, 0.6) plus a uniform (0,1) offset. A single sample was generated for each set of parameters and the parameters estimated. Figures 4, 5 and 6 are comparative box and whisker plot with the same layout as Figures 1-3, showing the relationship between the true parameter values and the estimates. Formal calculations confirm what is obvious in these figures; there is no perceptible bias in the estimates of either $\alpha$ or $\beta$: the lowess smooth can not be distinguished from the line of identity. There is a perceptible but small bias in $\lambda$. Each sample fit took 0.1 seconds. The model and algorithm therefore pass muster. Repeating the simulation using 0, 2, 3 and 4 refinement steps led to worse estimates of $\lambda$ and essentially the same $\alpha$, $\beta$ confirming one refinement step as the sweet spot.

**Residuals**

The $n$x$I$ matrix of residuals is defined by

$$e_{ji} = X_{ji} - \alpha_i - \beta_i \widehat{\mu}_i$$

The variance of their model counterpart with $\mu_i$ known is $\lambda_i \kappa^2 \{\rho^2 + (\alpha_i + \beta_i \mu_j)^2\}$. Substituting the estimates of these parameters and ignoring the second-order effects of their random variability leads to the scaled residuals

$$r_{ji} = \frac{X_{ji} - \widehat{\alpha_i} - \widehat{\beta_i}\widehat{\mu_j}}{\widehat{\kappa_i}\sqrt{\widehat{\lambda_i}\{\widehat{\rho^2} + (\widehat{\alpha_i} + \widehat{\beta_i}\widehat{\mu_j})^2\}}} \qquad (7)$$

which, under the model, are approximately standard normal and correlated between instruments, but not across samples.

The residual analysis tools of conventional regression (4, 5) can be applied to these scaled residuals. In particular a scatter plot against the estimated $\mu$, and one against their ordering by $\mu$ (an "index" plot) are informative. Under the model, these plots should have constant vertical height, or the precision profile is not the function assumed, and should have no trend, or the model of linearity is wrong. It is helpful to add a lowess smooth (5) as a visual linearity cue. A quantile-quantile (QQ) plot is the other standard graphic for checking whether the residuals follow the assumed normal distribution.

**Outliers**

Outliers may be visible in any of these plots. Because an outlier on one of the assays distorts all $I$ of that sample's residuals, finding outliers is necessarily a multivariate exercise using the vector of residuals of the sample from the different assays. The starting point for the search is to fit the multi-instrument model, calculate the $n$x$I$ matrix of residuals, and use the fitted precision profile model to transform them to the putatively N(0,1) scale.

Write R for the $n$x$I$ matrix of scaled residuals and $r_i$ for its $i^{th}$ row, corresponding to the $I$ scaled residuals of the $i^{th}$ sample. Find the squared Mahalanobis distances of each sample (6) from the overall mean: write $\bar{r}$ for the $I$ component mean vector and $S$ for their $I$x$I$ covariance matrix. The Mahalanobis distances of the samples from the mean, which show the relative atypicality of each sample, are defined by

$$D_i = (r_i - \bar{r})^T S^{-1} (r_i - \bar{r})$$

The $D_i$ may be transformed (7) to Hotelling $T^2$ statistics which have the virtue of greater familiarity, being the maximized square of a two-sample t statistic between that sample and the rest of the data set. As a practical matter, the residuals, all being based on deviations from that sample's consensus, are highly correlated, so the covariance matrix S as defined here is inevitably singular or close to singular. For this reason, in calculating the multivariate diagnostics, we drop one column of R and work with the remaining $I$-1. It is immaterial which column is dropped.

*Stage 1: Forward selection*

The two-stage outlier identification procedure follows Rosner's ESD model (8). Decide K, the maximum number of outliers to be searched for. The value of $K$ is not critical other than it must be large enough to accommodate the actual number of outliers in the data. A reasonable default is 5% of the sample size.

Find the sample with the largest $D_i$. Put this sample on a list of suspect cases, and remove it from the sample. Repeat the entire calculation with the remaining $n$-1 cases, recalculating the mean vector and covariance matrix of the new scaled residuals. Find the sample amongst these cases with the largest $D_i$ and add it to the list of suspects, removing it from the sample. Continue in this way until $K$ samples have been moved from the "clean" to the suspect list.

*Stage 2: Reinclusion*

After this initial sorting, we are left with $n$-$K$ putatively clean cases. Fit the model to these clean cases, finding their estimated $\alpha, \beta, \lambda, \sigma, \kappa$ values, and use these values to find the residuals and scaled residuals of the $n$-$K$ cases.

Using these parameter estimates and the estimated $\mu$ values from the full-sample fit, calculate the $K$x$I$ matrix of the residuals of the suspect cases, and rescale them to N(0,1). Write $r_i$ for the $i^{th}$ vector of these scaled residuals of the held-out cases and $\bar{r}_i$ and $S_i$ for their mean vector and covariance matrix.

Compute $T_i = \frac{N-K}{N-K+1}(r_i - \bar{r}_i)^T S_i^{-1}(r_i - \bar{r}_i)$. This is a Hotelling $T^2$ test statistic for the two-sample comparison of the held-out case with the clean samples. It is a standard multivariate outlier test statistic (7 section 8.1).

Find the smallest of these $T_i$ and calculate its P value using the known distribution of Hotelling $T^2$. Convert to a Bonferroni outlier P value by multiplying it by $n$-$K$+1. If this Bonferroni P value exceeds the cutoff selected for significance, then conclude that this suspect case is not a significant outlier, move it back to the clean cases, refit the model to the updated set of clean cases, and repeat the reinclusion test on the remaining suspects.

Continue in this way until you either reinclude all the suspects and conclude that there are no significant outliers, or find a significant P in which case you conclude that all cases still on the suspect list are significant outliers.

*Followup*

The analysis finds outlying vectors of scaled residuals. Commonly one element of $r$ of a case will be numerically much bigger than the others, indicating that the corresponding instrument produced a questionable value, with the others pushed in the opposite direction to compensate.

**Example**

Some of these issues can be illustrated by a simulated example: 120 samples with true values $\mu$ in geometric progression from 8 to 80 were generated and measured using four instruments. The precision profile parameters were $\sigma$=3, $\kappa$=0.08 from which $\rho$=3/0.08 = 37.5. The characteristics of the instruments were:

| Instrument | 1 | 2 | 3 | 4 |
|---|---|---|---|---|
| $\alpha$ | -1 | 1 | 2 | -2 |
| $\beta$ | 1.1 | 0.9 | 1.2 | 0.8 |
| $\lambda$ | 1.8 | 0.45 | 0.9 | 0.9 |

Figure 7 plots the individual assays against the consensus fitted $\mu$. The dotted line is the line of identity, and the dashed line the fitted regression. The fitted model is

| | | | | |
|---|---|---|---|---|
| $\sigma$ | 1.8117 | $\kappa$ | 0.0790 | |
| | Inst_1 | Inst_2 | Inst_3 | Inst_4 |
| $\alpha$ | -1.5860 | 1.1220 | 2.2050 | -1.7400 |
| se | 0.5992 | 0.3031 | 0.5333 | 0.3425 |
| $\beta$ | 1.1413 | 0.8846 | 1.1849 | 0.7892 |
| se | 0.0235 | 0.0127 | 0.0192 | 0.0154 |
| $\lambda$ | 1.9440 | 0.3610 | 0.8710 | 0.8240 |
| se | 0.1690 | 0.0890 | 0.1530 | 0.1740 |

The estimates of $\alpha, \beta, \sigma, \kappa$ and $\lambda$ agree well with the true values.

Figure 8 is a plot of the scaled residuals versus the estimated $\mu$, and Figure 9 is an index plot of the scaled residuals. The dashed line is the lowess smooth, which hugs the axis in all the plots. The plots give no hint of changes in the vertical spread or curvature. This supports the assumed precision profile model and linearity. All four instruments pass QQ plot test for normality.

To illustrate outliers, the instrument 4 assay of one of the low samples was increased, and the instrument 2 assay of one of the high samples was decreased. Figure 10 is a scatter plot of the resulting sample with the two induced outliers shown as filled boxes. The fitted regression was

| $\sigma$ | 1.9186 | $\kappa$ | 0.0812 | |
|---|---|---|---|---|
| | Inst_1 | Inst_2 | Inst_3 | Inst_4 |
| $\alpha$ | -1.7200 | 1.2780 | 2.0310 | -1.5890 |
| se | 0.6081 | 0.3304 | 0.5410 | 0.4277 |
| $\beta$ | 1.1464 | 0.8737 | 1.1921 | 0.7878 |
| se | 0.0231 | 0.0149 | 0.0196 | 0.0171 |
| $\lambda$ | 1.7190 | 0.4200 | 0.7600 | 1.1010 |
| se | 0.2170 | 0.1320 | 0.1810 | 0.2940 |

The data set was run through the outlier identification procedure with K=4, giving

Forward selection

| case | Mahal D | Hotel T |
|---|---|---|
| 1 | 22.524 | 27.835 |
| 120 | 20.951 | 25.519 |
| 115 | 11.196 | 12.391 |
| 108 | 10.007 | 10.960 |

Backward reinclusion

| case | Bonf P |
|---|---|
| 108 | 1.0353 |
| 115 | 0.6610 |
| 120 | 0.0011 |

leading to the significant outliers

| case | Bon P | Inst_1 | Inst_2 | Inst_3 | Inst_4 |
|---|---|---|---|---|---|
| 1 | 0.000687 | -2.189 | -0.967 | -0.509 | 5.749 |
| 120 | 0.001138 | 0.017 | -4.958 | 1.474 | 3.922 |

The *r* values point to instrument 4 on sample 1 and instrument 2 on sample 120 with induced compensating scaled residuals on the other three instruments. The analysis has therefore correctly diagnosed the two induced outliers.

A real-world example comes from (3). Two preparation methods A and B were tested using instruments X and Y with three reagent lots being used for each combination. To keep the displays compact, we use two of the three reagent lots, leading to 8 "instruments". Each was used to assay 83 samples. Figure 11 plots the data against the fitted $\mu$. The fitted model was:

| $\sigma$ | 0.4469 | $\kappa$ | 0.0050 | | | | | |
|---|---|---|---|---|---|---|---|---|
| | A_X_1 | A_X_2 | A_Y_5 | A_Y_6 | B_X_1 | B_X_2 | B_Y_5 | B_Y_6 |
| $\alpha$ | -1.5950 | -0.8795 | -0.2959 | 0.1387 | 0.3488 | -0.8956 | 0.9801 | 2.1980 |
| se | 0.3408 | 0.2094 | 0.2439 | 0.1850 | 0.2540 | 0.2251 | 0.2073 | 0.2308 |
| $\beta$ | 1.0387 | 1.0168 | 1.0068 | 0.9997 | 0.9949 | 1.0054 | 0.9756 | 0.9620 |
| se | 0.0055 | 0.0037 | 0.0046 | 0.0034 | 0.0044 | 0.0041 | 0.0037 | 0.0040 |
| $\lambda$ | 1.8220 | 0.6480 | 0.6450 | 0.9320 | 1.0630 | 0.8580 | 0.8010 | 1.2300 |
| se | 0.8900 | 0.1820 | 0.1310 | 0.2620 | 0.2190 | 0.1650 | 0.1830 | 0.2260 |

The slopes and intercepts are clearly non-identical and equality of the $\lambda$ is questionable. Figure 12 is a scatter plot of each instrument's scaled residuals against the consensus. The A_X_1 plot show a problem – a visual outlier, shown as a solid box; this sample is also shown as a box in the other plots. A formal outlier screening led to the endpoint:

| case | Bon_P | A_X_1 | A_X_2 | A_Y_5 | A_Y_6 | B_X_1 | B_X_2 | B_Y_5 | B_Y_6 |
|---|---|---|---|---|---|---|---|---|---|
| 7 | 1.8897e-13 | -10.623 | -0.869 | 2.285 | -1.082 | -2.897 | 2.846 | 2.854 | 3.386 |

The case is a massive outlier. The residuals point to A_X_1 as the culprit. Refitting the model excluding this case gives:

| $\sigma$ | 0.3561 | $\kappa$ | 0.0058 | | | | | |
|---|---|---|---|---|---|---|---|---|
| | A_X_1 | A_X_2 | A_Y_5 | A_Y_6 | B_X_1 | B_X_2 | B_Y_5 | B_Y_6 |
| $\alpha$ | -1.3520 | -0.9160 | -0.3841 | 0.1445 | 0.4495 | -0.9645 | 0.9109 | 2.1120 |
| se | 0.2199 | 0.2157 | 0.2222 | 0.1867 | 0.2460 | 0.2018 | 0.1919 | 0.2118 |
| $\beta$ | 1.0354 | 1.0174 | 1.0082 | 0.9997 | 0.9934 | 1.0063 | 0.9765 | 0.9631 |
| se | 0.0039 | 0.0038 | 0.0043 | 0.0034 | 0.0043 | 0.0038 | 0.0034 | 0.0038 |
| $\lambda$ | 0.9560 | 0.7790 | 0.7030 | 1.1300 | 1.2030 | 0.9420 | 0.9190 | 1.3680 |
| se | 0.1970 | 0.1580 | 0.1380 | 0.1880 | 0.2070 | 0.1750 | 0.1760 | 0.2020 |

The slopes barely change with the outlier deletion and only the first intercept is affected appreciably. However $\lambda_1$ and its standard error drop dramatically, showing that the outlier had a substantial effect on the estimated variability of A_X_1. With the deletion, all $\lambda$ are about the same, indicating a common precision profile for the 8 "instruments". The plots of the cleaned data set were unremarkable suggesting that, apart from the outlier reading, the model was a good fit to the data set.

The object of the experiment was to establish the effects of method of preparation and instrument. Analyzing the 8 intercepts and slopes with a two-way analysis of variance for the factors B vs A and Y vs X gives:

| Intercept | coeff | std_err |
|---|---|---|
| (Intercept) | -1.323 | 0.4099 |
| AB | 1.254 | 0.4733 |
| XY | 1.391 | 0.4733 |

Slopes

| Term | coeff | std_err |
|---|---|---|
| (Intercept) | 1.028 | 0.0055 |
| AB | -0.0303 | 0.0063 |
| XY | -0.0263 | 0.0063 |

One should not take the standard errors too literally as the 8 coefficient estimates are highly correlated. Nevertheless the fits suggest that B as against A, and Y as against X each reduce the slope by about 0.03, leading to a concomitant increase of about 1.3 in the intercept.

We repeated the analysis using all the lots and reached essentially the same conclusion, which also agrees with that from a multi-instrument PB as reported in (3).

## Conclusion

Multiple instruments measuring the same samples can be compared pairwise, but unless one of the methods is enshrined as the comparator, a more satisfying approach is to find a consensus among them and compare each instrument to the consensus, reducing the number of comparisons and the likelihood of anomalous or inconsistent conclusions. We provide a parametric approach for doing so when a model of normally-distributed assay random errors with mutually proportional precision profile models fits. Apart from the fitting, we sketch checks for model fit, and a formal outlier analysis. If this parametric model does not fit, then the multi-instrument Passing Bablok approach can be used. The methods discussed here are supported in the CRAN package ppwdeming (9) procedures multi_PWD for fitting the model, multi_PWD_inf for inference using the jackknife, and multi_PWD_out for outlier screening. The two-instrument residual analysis tools in the package apply also to the multi-instrument setting.

One feature/limitation of the formulation may not be obvious – whereas the methodology imposes no obvious limitations on the $\alpha$ and $\beta$ it assumes implicitly that the $\beta$ are all positive, without which the latent $\mu$ are conceptually elusive and possibly hard to estimate. This is not true of the two-instrument formulation, whose $\mu$ is tied to $X$, and where the sign of $\beta$ is immaterial. This limitation is, however, not relevant in methods comparison studies.

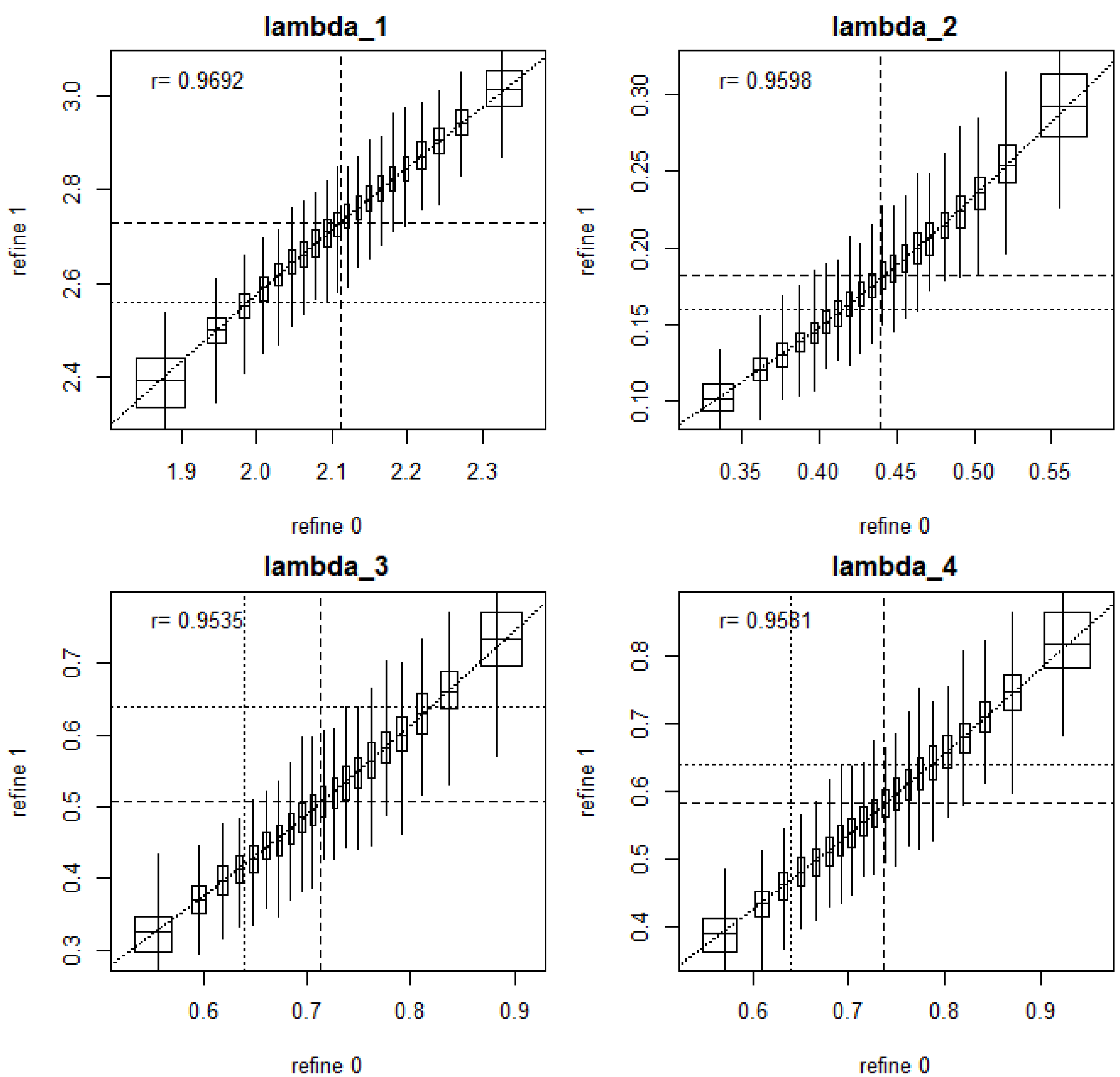

Figure 1 Four instruments' estimated $\lambda$.

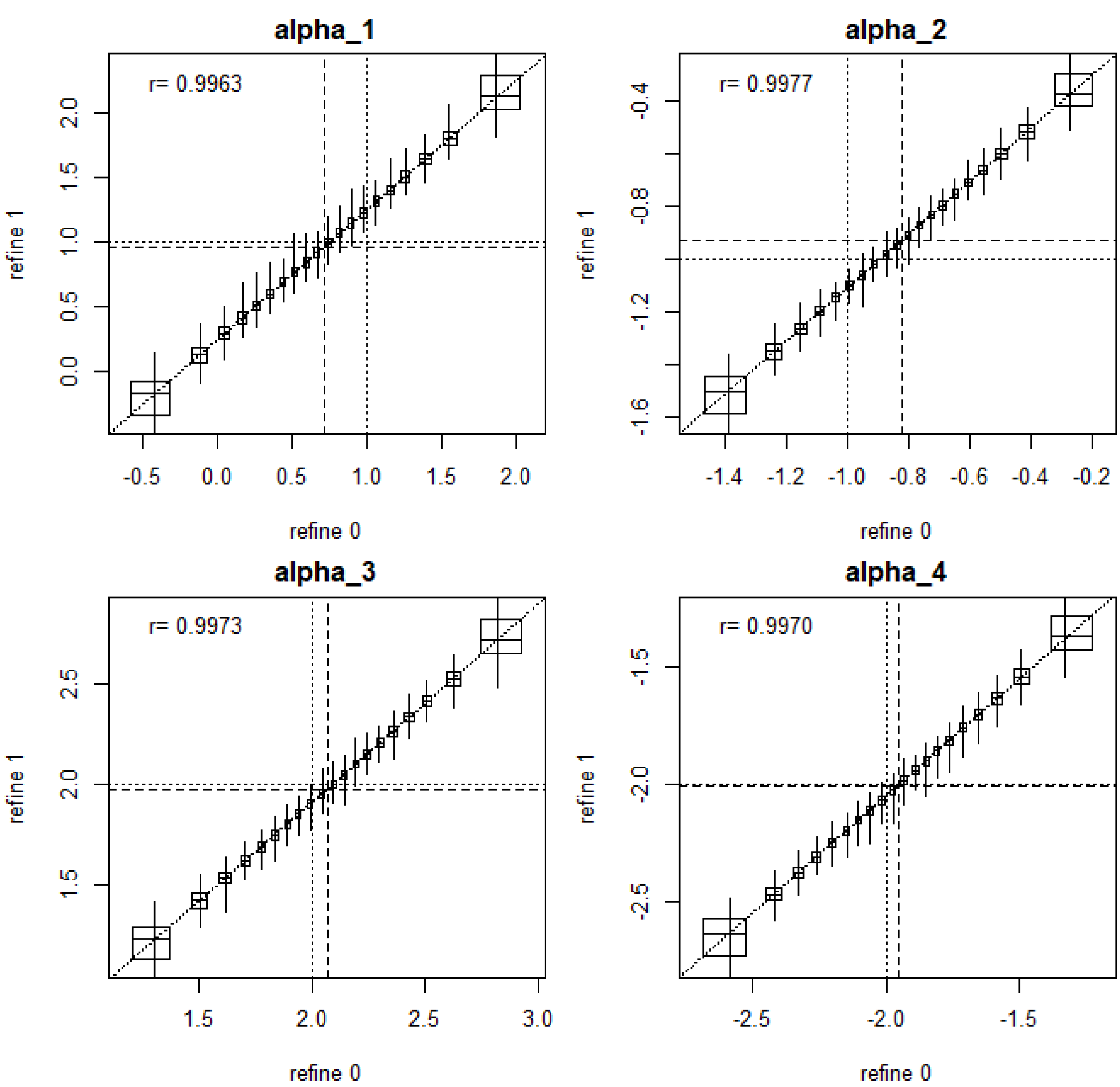

Figure 2 Four instruments' estimated $\alpha$.

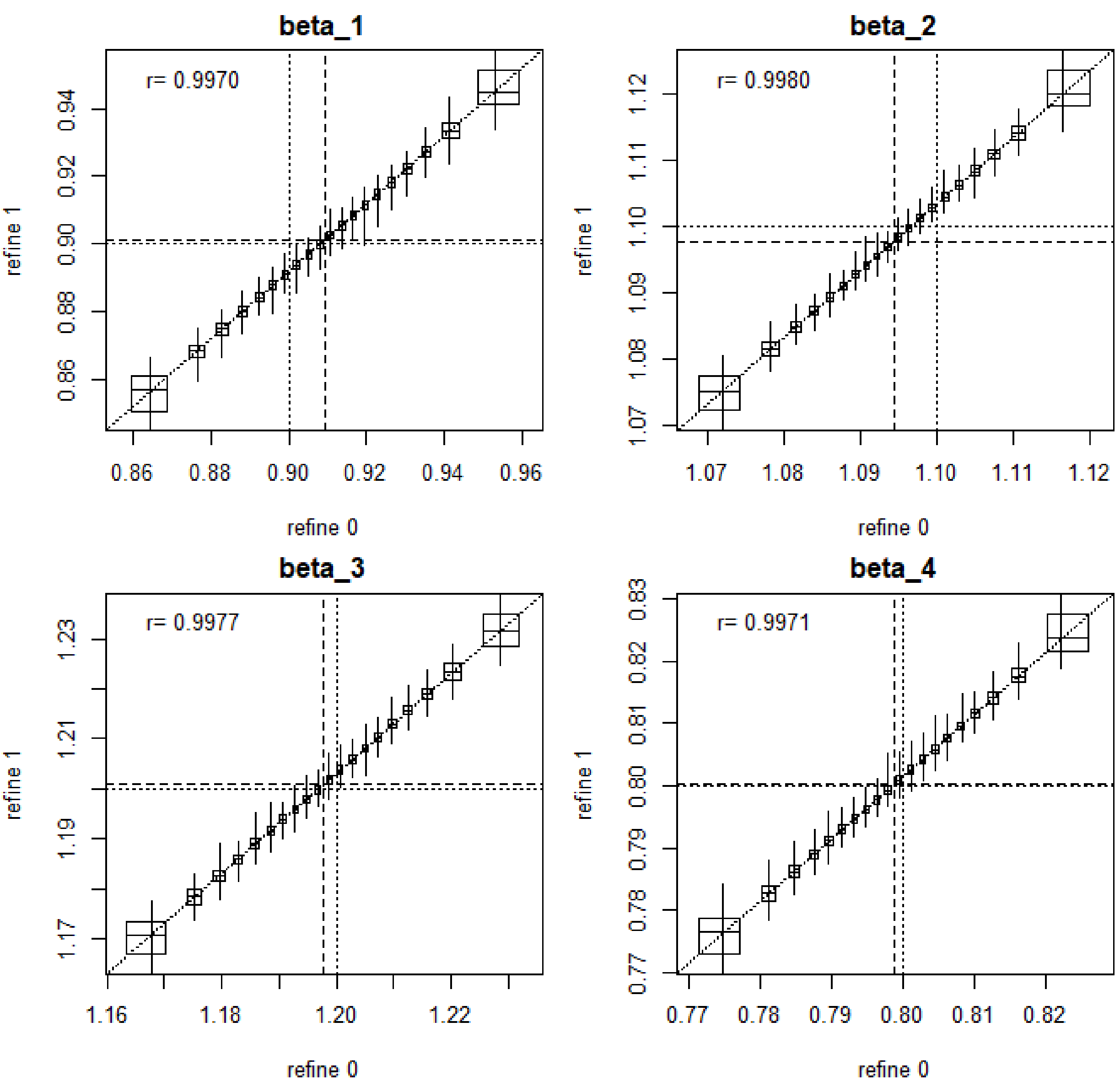

Figure 3. Four instruments' estimated $\beta$.

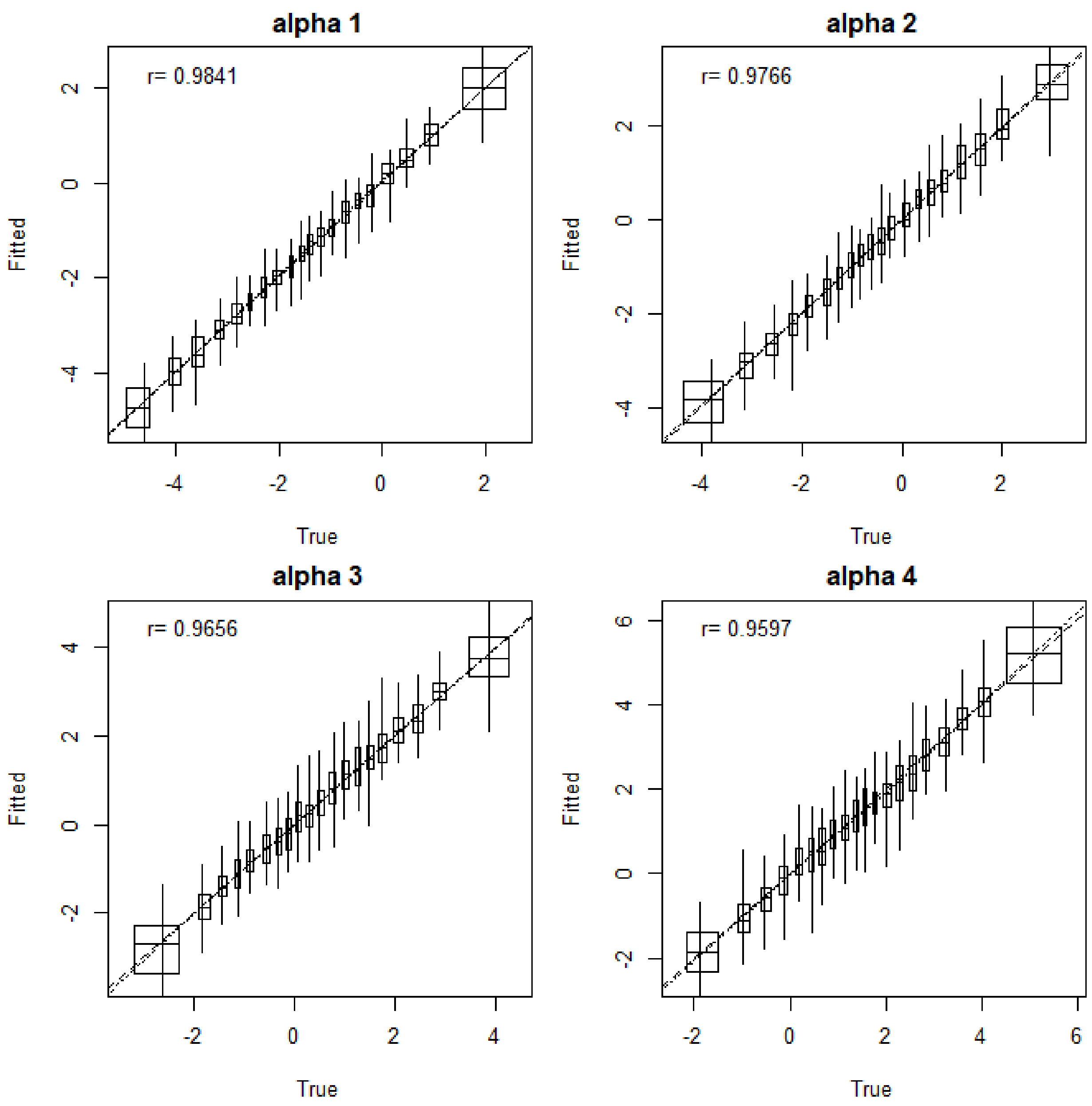

Figure 4. Broader simulation $\alpha$

\

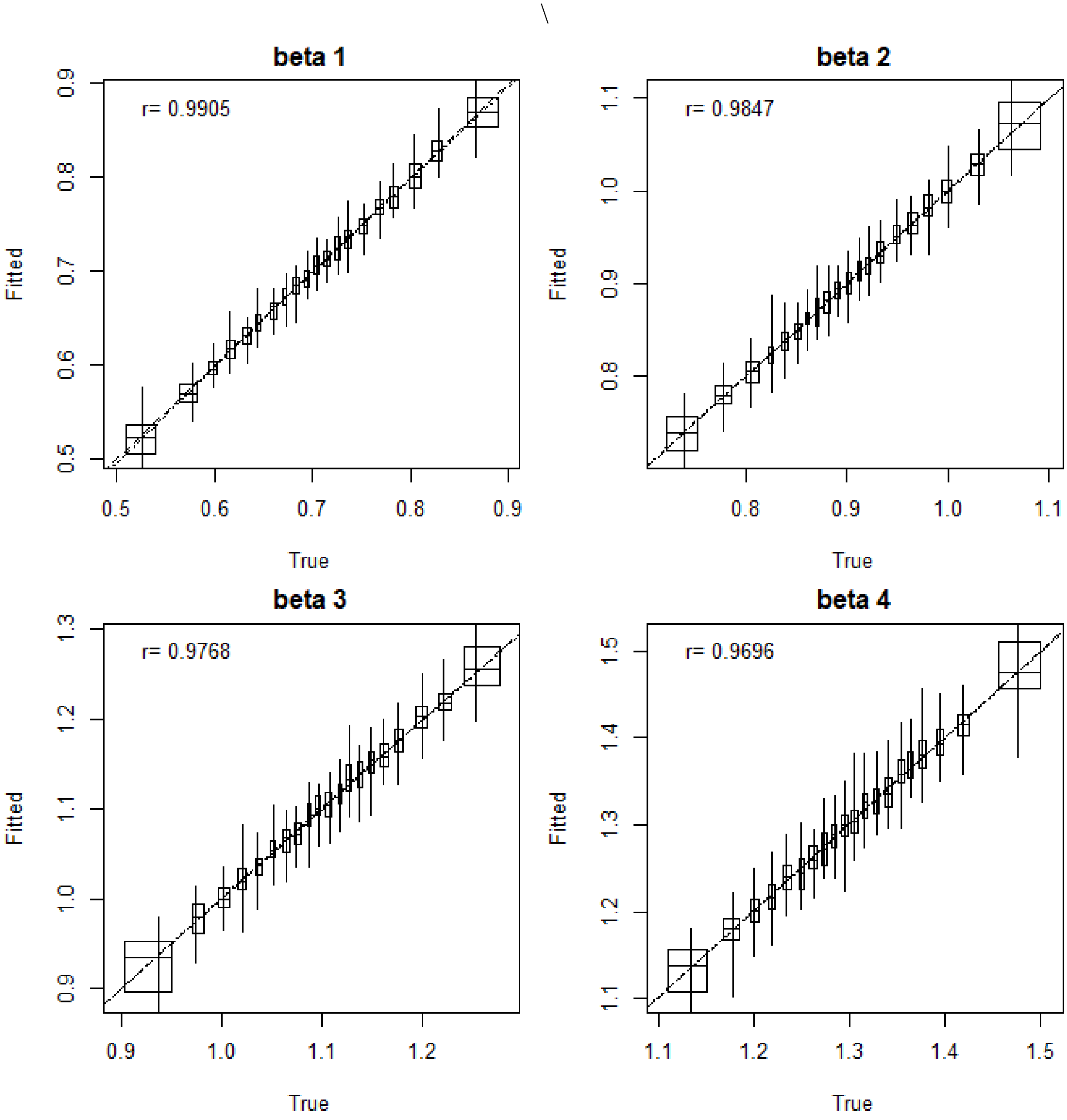

Figure 5. Broader simulation $\beta$

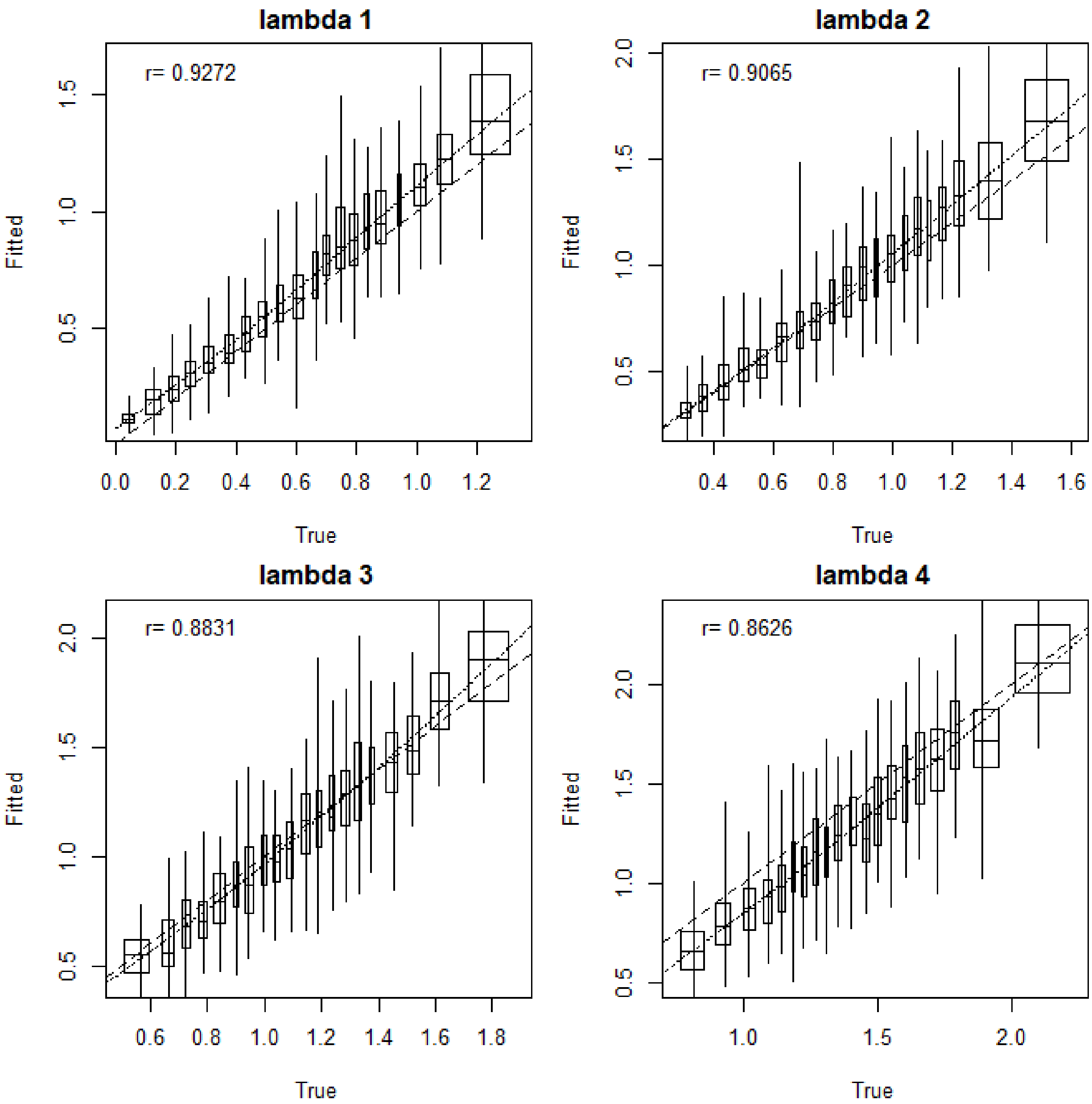

Figure 6. Broader simulation $\lambda$

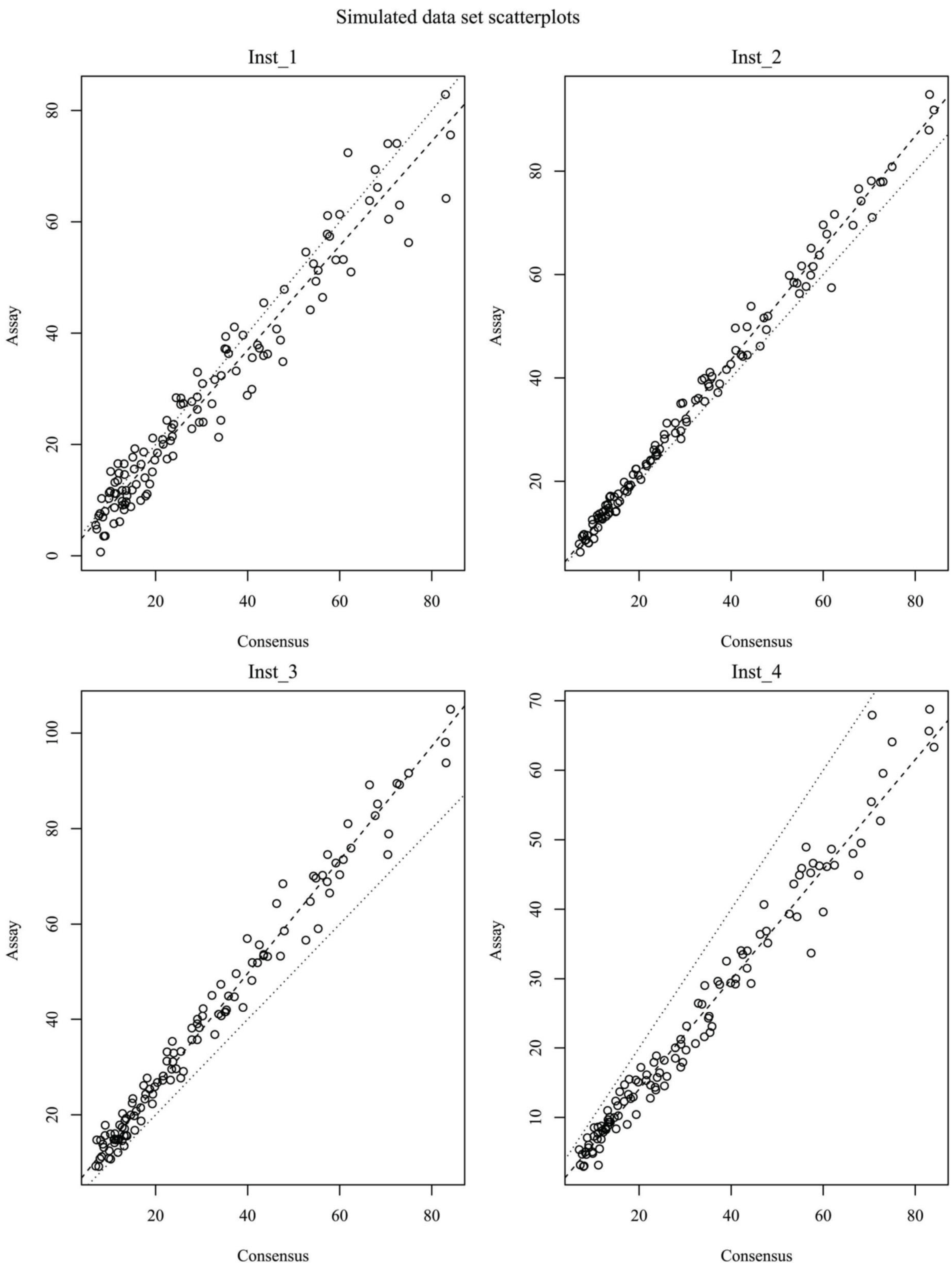

Figure 7. Assays vs consensus, example 1

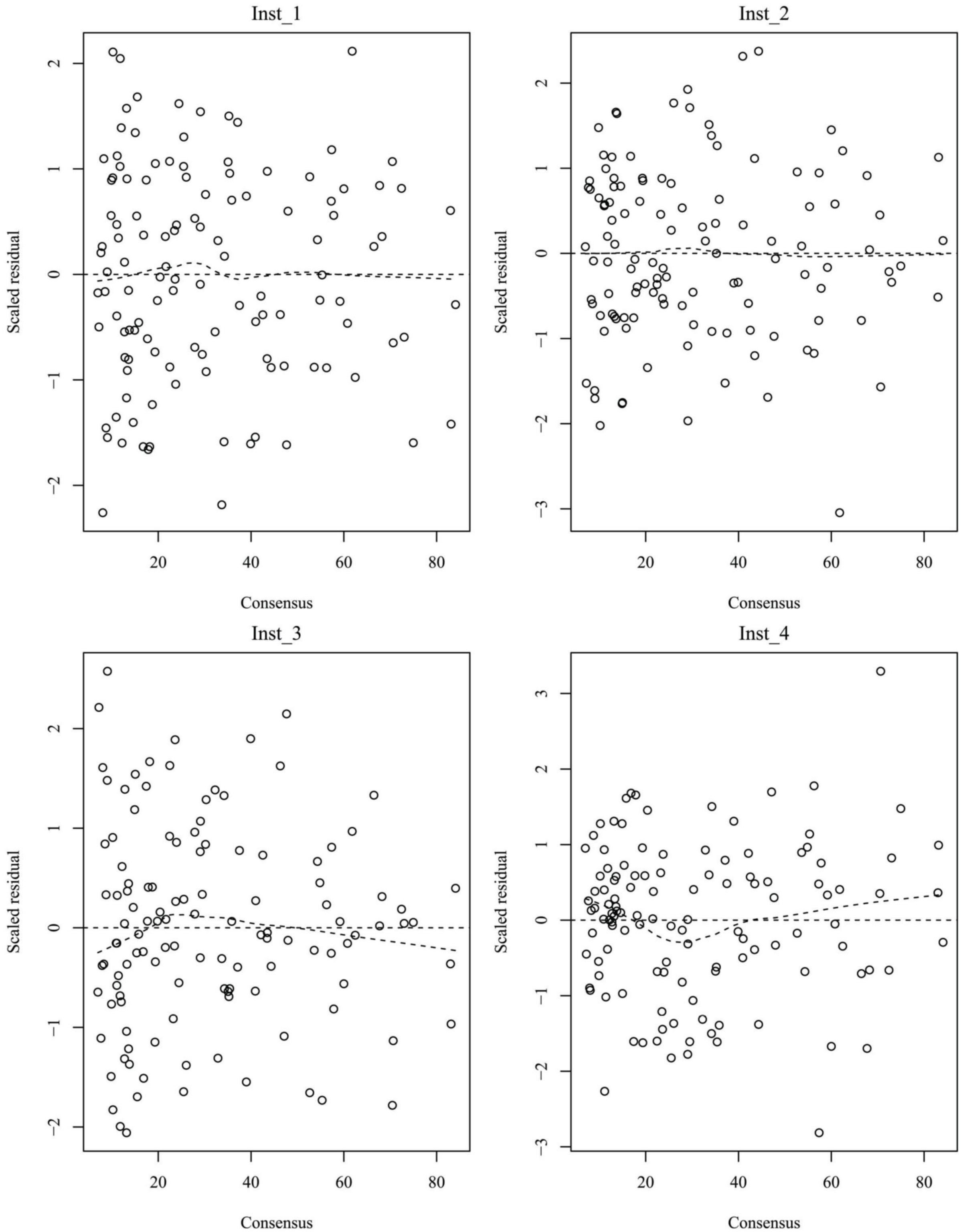

Figure 8. Scaled residuals vs consensus example 1

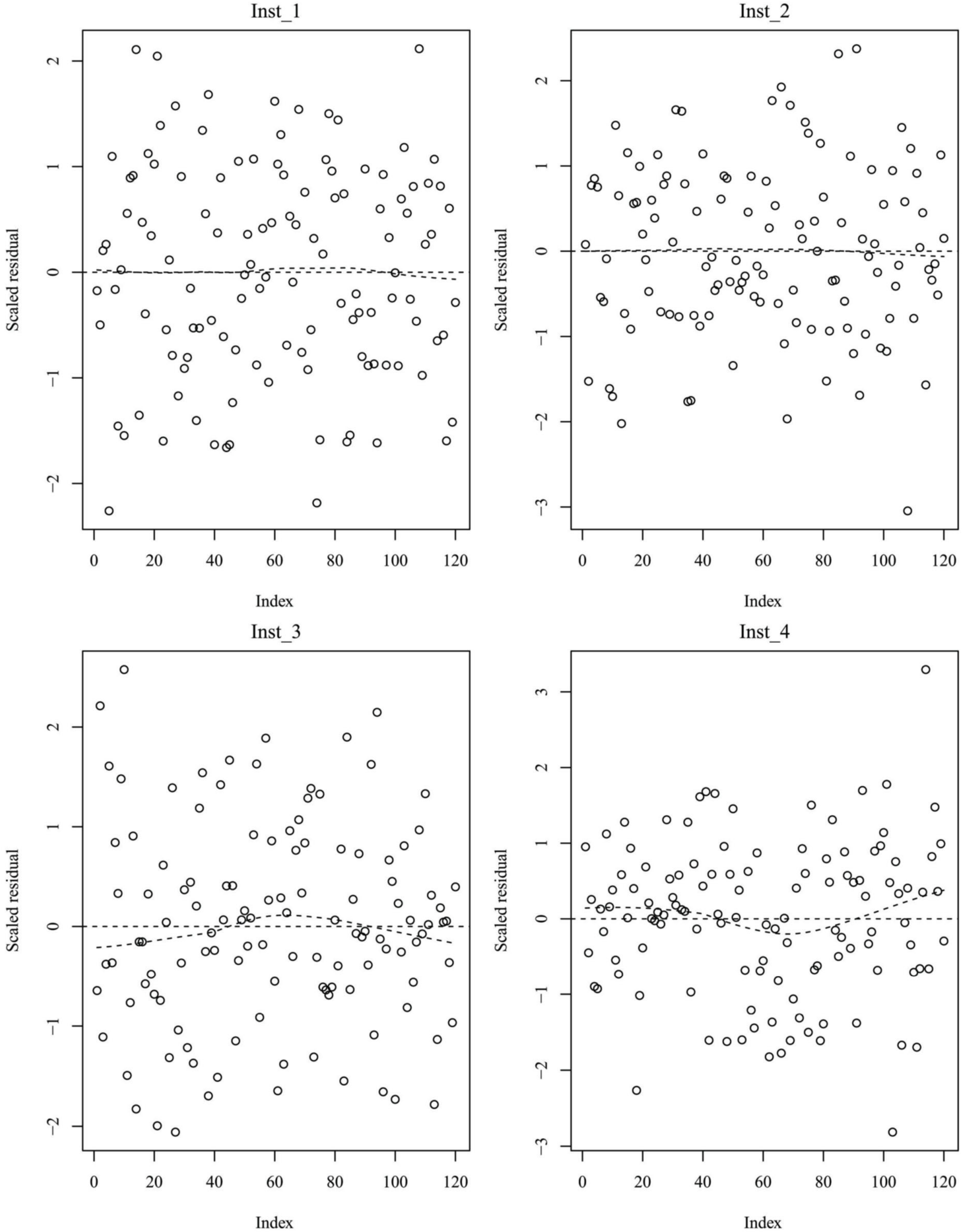

Figure 9. Index plot of scaled residuals, example 1

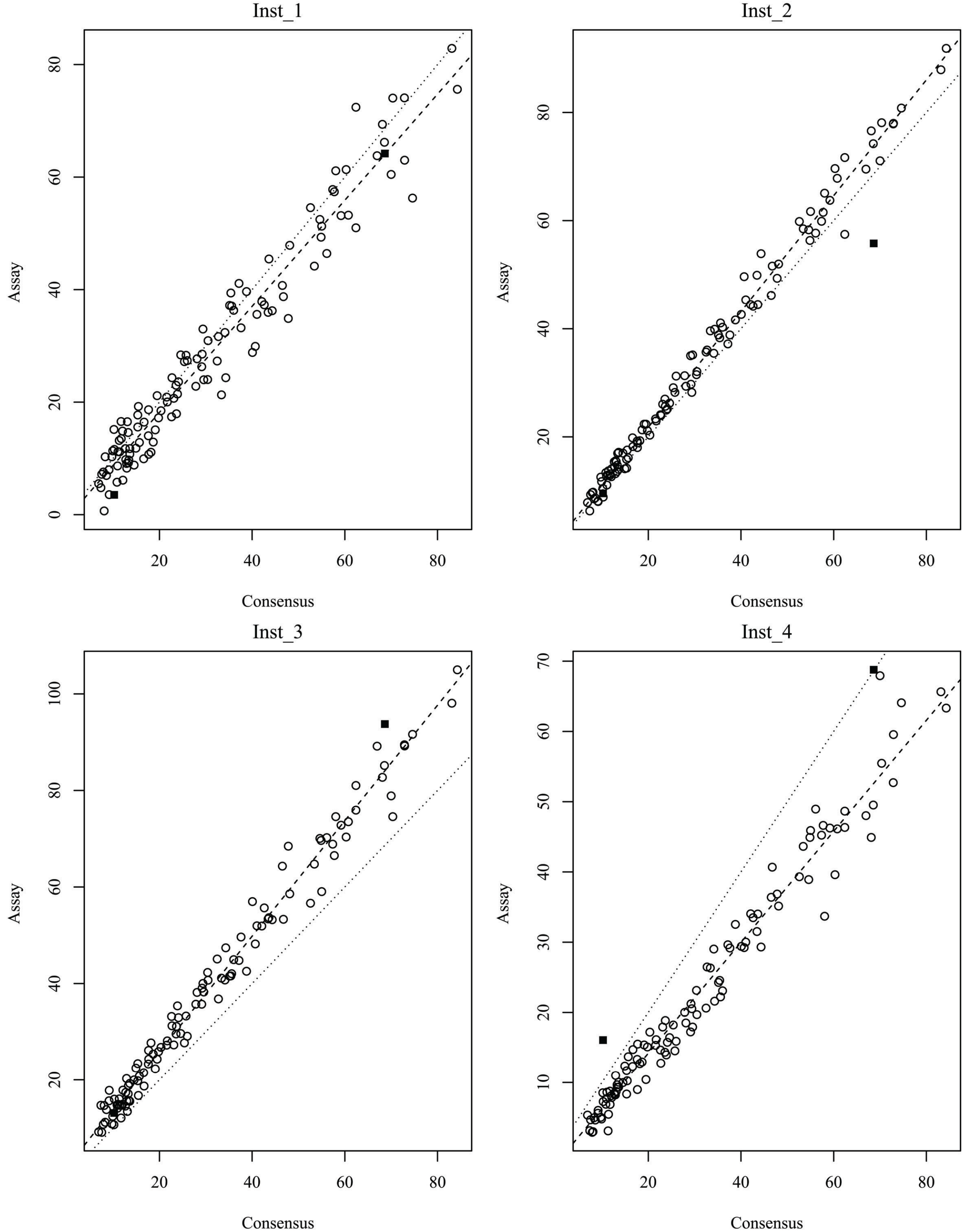

Figure 10. Two outliers added

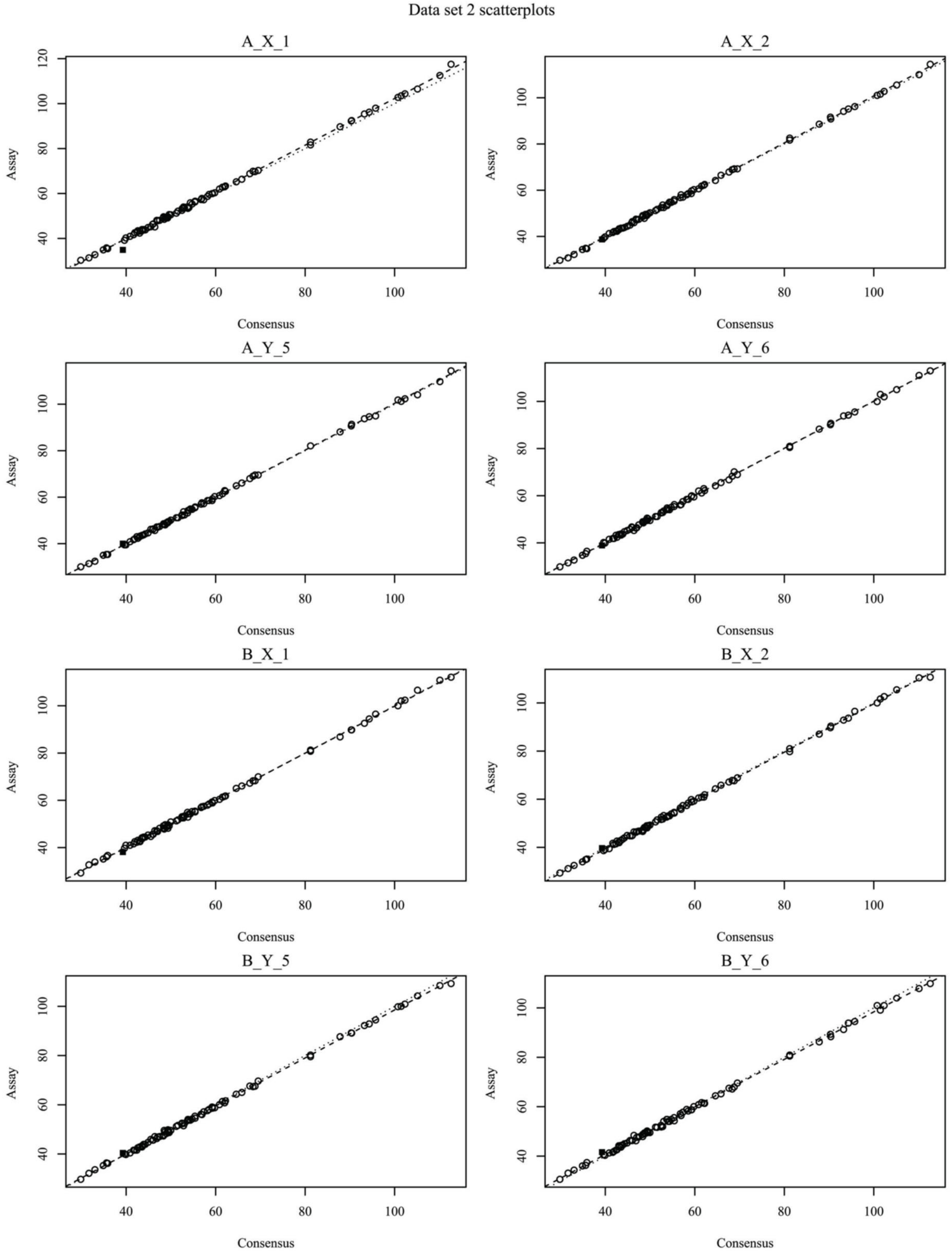

Figure 11. Second example scatter plot

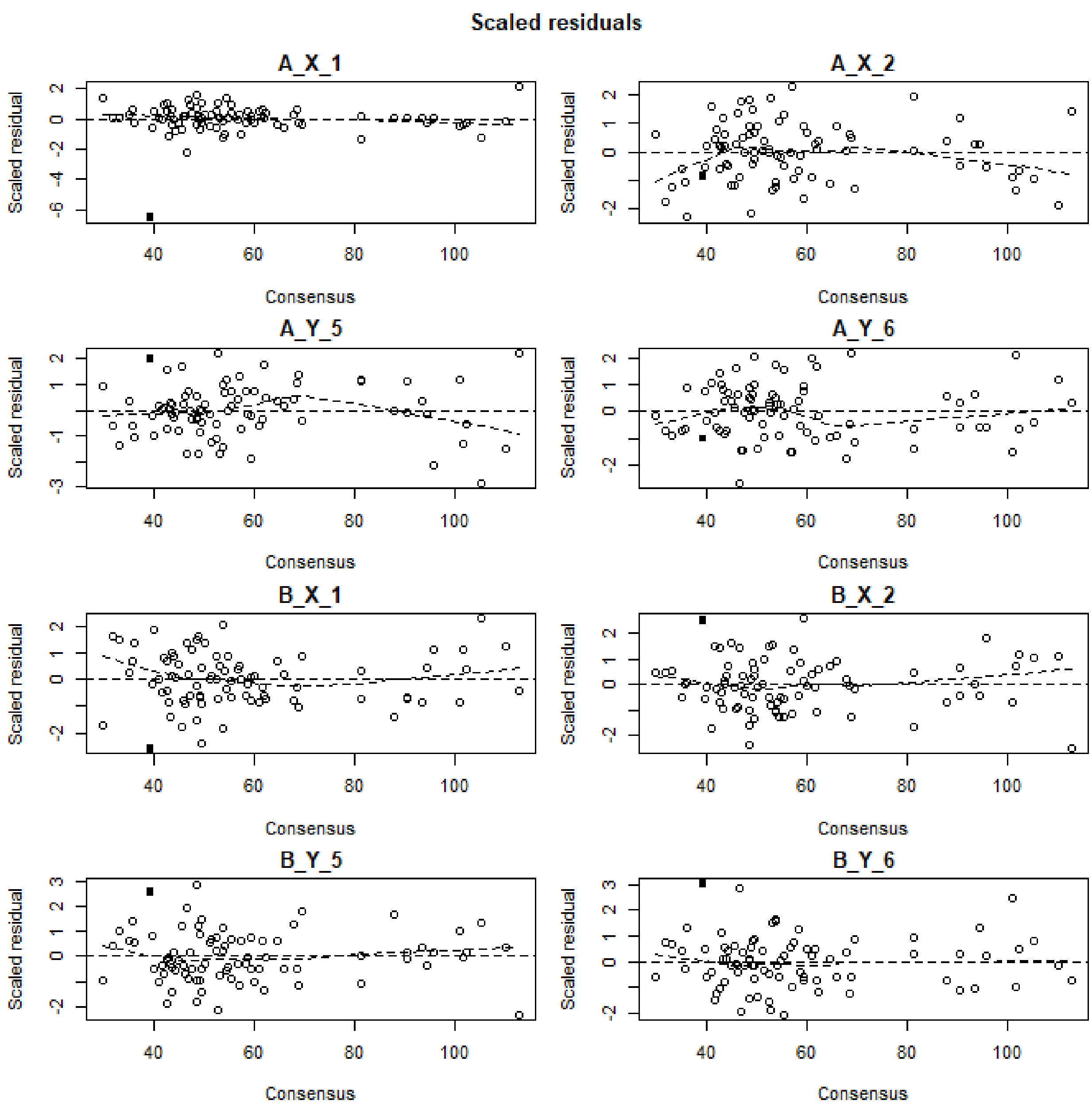

Figure 12. Scaled residuals, second example